\RequirePackage{fix-cm}
\documentclass[twocolumn,epjc3]{svjour3}  
\smartqed  
\RequirePackage{graphicx}
\usepackage{amssymb}
\usepackage{lipsum}
\usepackage{subcaption}
\usepackage{graphicx}
\usepackage{dcolumn}
\usepackage{bm}
\usepackage{epsfig,amsmath}
\usepackage{amssymb}
\usepackage{float}
\usepackage{xcolor}
\usepackage{multirow}
\usepackage{gensymb}
\usepackage{ulem}
\usepackage[pagebackref=false, colorlinks=true]{hyperref}
\definecolor{redish}{rgb}{0.7,0.2,0.0}  
\definecolor{bluish}{rgb}{0.2,0.5,0.8}

\hypersetup{linkcolor=redish,          
	citecolor=blue,        
	filecolor=magenta,      
	urlcolor=bluish}          

\DeclareFontFamily{U}{rsfs}{}         
\DeclareFontShape{U}{rsfs}{m}{n}{<5> rsfs5 <6><7> rsfs7          %
	<8><9><10><10.95><12><14.4><17.28><20.74><24.88> rsfs10}{}

\journalname{Eur. Phys. J. C}
\begin{document}

\title{Magnetically arrested transmutation of a compact star}

\author{H. A. Adarsha\thanksref{e1,addr1}
        \and
        Chandrachur Chakraborty\thanksref{e2,addr1}
        \and
       Sudip Bhattacharyya\thanksref{e3,addr2}
}

\thankstext{e1}{e-mail: adarsha.mcnsmpl2023@learner.manipal.edu}
\thankstext{e2}{e-mail: chandrachur.c@manipal.edu}
\thankstext{e3}{e-mail: sudip@tifr.res.in}

\institute{Manipal Centre for Natural Sciences, Manipal Academy of Higher Education, Manipal 576104, India \label{addr1}
           \and
          Department of Astronomy and Astrophysics,
Tata Institute of Fundamental Research, Mumbai 400005, India \label{addr2}
}

\date{Received: date / Accepted: date}

\maketitle

\begin{abstract}
We introduce a novel mechanism---Magnetically Arrested Transmutation (MAT)---which could be a viable model to account for the observed over-representation of {\it magnetic} white dwarfs (WDs) near the Galactic centre (GC), and the presence of a magnetar as opposed to the absence of ordinary pulsars in the same region.
In this scenario, compact stars accumulate asymmetric or non-self-annihilating dark matter particles, eventually forming an endoparasitic black hole (EBH) of initial mass $M_0$ at their core. 
Although such EBHs generally grow by accreting host matter, we show that sufficiently strong core magnetic fields can establish pressure equilibrium, thereby stalling further accretion and halting the star’s transmutation into a black hole. 
We derive the conditions for this MAT to occur, identifying a critical parameter $\beta$, that encapsulates the interplay between the magnetic field strength, host matter density, and EBH mass. 
For $0 < \beta \leq 4/27$, the growth of the EBH is arrested, limiting its final mass ($M_{\rm f}$) to $M_0 <M_{\rm f} \leq 3/2M_0$, whereas for $\beta > 4/27$, full transmutation may ensue. 
We argue that highly magnetized WDs may survive near the GC due to the MAT mechanism, as do high-spin ordinary WDs, despite hosting a central EBH.
We also speculate a possibility that the magnetar PSR J1745-2900 survives near the GC due to the MAT mechanism. Overall, the MAT framework may explain an elevated population of magnetic WDs in dense dark matter environments, and hence could be tested and should have implications for understanding dark matter and compact objects.

\end{abstract}

\section{Introduction}
\label{intro}

Compact stars with high magnetic fields are rich laboratories of fundamental physics as well as important objects to understand stellar evolution. 
Based on theoretical knowledge of stellar evolution, stellar dynamics, and the observed density of stars near the Galactic centre (GC), one could predict the mass distribution and number density of neutron stars (NSs) and white dwarfs (WDs) around the GC, which is densely populated with stars. In this region, one observes more magnetic cataclysmic variables or simply WDs with high magnetic fields than expected \cite{Yu_2022}.

WDs with high surface magnetic fields (up to $\sim 10^8$ G) are known to exist in the GC or the nuclear star cluster (NSC), as a part of magnetic cataclysmic variables (mCV) which are semi-detached binaries with a WD accreting matter from a main sequence or sub-giant companion star \cite{Xu_2019}. 
The mass of these WDs is usually found to be above the average mass $0.6 M_\odot$ and are known to be slow rotating (see, for example \cite{Vermette_2023}). 
On the other hand, to explain over-luminous Type Ia supernovae, the existence of magnetic white dwarfs (mWDs) has also been proposed by \cite{Kalita2019,Bhattacharya_2022}. 
One finds that these kinds of WDs (also in mCVs) could be found more predominantly in the GC \cite{ferrario_magnetic_2015,Yu_2022}. 
While there has been debate over the origin of mWDs, the observed higher fraction of mCVs in the GC as opposed to a lower fraction of the non-magnetic or ordinary CVs \cite{Yu_2022} compared to the predictions is not satisfactorily explained. 
Although there have been detections of accreting, ordinary WDs known as dwarf novae (DN) \cite{Yu_2022}, their numbers are lower compared to the number of mCVs. 
While this could simply be due to the bias in observation \cite{Pretorius_2013} it is interesting to explore alternate explanations for a real higher fraction of mCVs/mWDs in the vicinity of GC.

In case of NSs, the discrepancy is so stark that not a single pulsar is observed within a $\sim$10 pc radius of the central black hole Sgr A* when the expectation was to find about $10^3$ (or, at least 10) pulsars \cite{Pfahl_2004}. 
This is the ``missing pulsar problem" (MPP) \cite{Dexter_2014}.
The problem turned even trickier when a magnetar PSR J1745-2900 was discovered \cite{Mori_2013,Pearlman_2018} in the vicinity of Sgr A*, suggesting that the explanation invoking temporal scattering was not enough to explain the non-detection of ordinary pulsars in various surveys \cite{Johnston2006,Deneva_2009,Macquart_2010,Bates2011}. 
Several different solutions have been suggested for the problem. 
The collapse of young magnetars (considering them to be quark stars), due to a mini black hole formed at their centre by $\Gamma$ photon-quark matter interaction in a magnetic dual chiral density wave
(MDCDW) phase \cite{ferrer_axion-polaritons_2024} is one such solution. 
Here the authors assume that most pulsars formed in the GC are magnetars (not ordinary pulsars) like the PSR J1745-2900, that collapse under the axion-polaritons mechanism within a short span of time and hence, it is expected that we observe only very young magnetars like PSR J1745-2900, which will also collapse soon.
Another important suggestion is the collapse of pulsars due to transmutation by captured primordial black holes (PBH) or dark matter (DM) particles \cite{Fuller_PBH_r-process,fuller2015}, resulting in fast radio bursts (FRB), though the PBH-capture may not be enough to explain the MPP \cite{Caiozzo_2024}. 
In light of the above solution to MPP, the survival of the magnetar PSR J1745-2900 is surprising and demands an explanation.

In this letter, using the simple mechanism of formation of endoparasitic black hole (EBH) \cite{goldman_weakly_1989,kouvaris_constraining_2011,mccullough_capture_2010,mcdermott_constraints_2012,bramante_detecting_2014,acevedo_supernovae_2019,dasgupta_low_2021} inside mWDs and magnetars, we propose a solution for the observed higher concentration of mWDs near the GC where DM capture, formation of EBHs at the core of WDs and their subsequent transmutation were expected similar to NSs, and speculatively, the presence of a lone magnetar PSR J1745-2900 as well.
The mechanism consists of three parts: (1) asymmetric or non-self-annihilating DM capture by WDs and NSs, accumulating it in the core till the collapse criterion \cite{Chakraborty_2024,Chakraborty_low_mass_nakedsingularity2024} is satisfied leading to the formation of an EBH, (2) growth of EBH by accreting matter from its host and (3) magnetic field acting as a deterrent to the fast growth of the formed EBH thereby preventing the fast transmutation of mWDs and magnetars. 
We consider a scenario similar to the magnetically arrested disk (MAD) framework \cite{Narayan2003MAD}, where energy equipartition between infalling matter and magnetic fields leads to suppressed accretion \cite{Igumenshchev_2008}. Adapting this idea to nearly spherical accretion from a slowly rotating host, we demonstrate how strong magnetic fields can slow down or stall accretion onto EBH.
We also calculate the necessary core magnetic field for stalling accretion by EBH inside mWDs and magnetars.

\section{Magnetically arrested transmutation: Formalism}
\label{sec1}

Compact stars such as NSs and WDs residing in DM-dense environments, most notably in the GC, can efficiently accumulate non-self-annihilating or asymmetric DM particles. This occurs through repeated scattering interactions between DM particles and the nuclei/nucleons constituting the stellar material. The capture rate $F$ for asymmetric DM is given by  Eq. (3.1) of \cite{Chakraborty_low_mass_nakedsingularity2024}, or, equivalently by  Eq.~(12) of \cite{mcdermott_constraints_2012}. $F$ depends on the ambient DM density, the DM velocity dispersion ($\bar{v} \approx 220~\mathrm{km\,s^{-1}}$), the DM particle mass $(m_{\chi})$ and the DM--nucleon scattering cross section $\sigma$.

Once captured, the DM particles undergo repeated scatterings inside the star and eventually settle into an isothermal, approximately spherical distribution at the stellar centre \cite{mcdermott_constraints_2012}. When the accumulated number of DM particles satisfies the collapse criterion: $N \geq \max\left[ N_{\rm Ch},\, N_{\rm self} \right]$,
where $N$ is the total number of captured DM particles, 
$N_{\rm self}$ is the number required for the onset of self-gravitation, 
and $N_{\rm Ch}(m_{\chi})$ is the Chandrasekhar limit for DM of mass $m_{\chi}$, 
the DM core becomes gravitationally unstable. It then collapses to form a tiny BH of mass $M_0 = m_{\chi} N$ at the core of the host star. The detailed formation process of such an 
EBH inside a host star is discussed in Secs.~2 and 3 of 
\cite{Chakraborty_low_mass_nakedsingularity2024} and references therein.

When an EBH of mass $M_0$ forms inside a compact star by the accumulation of asymmetric DM particles, the EBH starts to consume matter from its near vicinity \cite{mcdermott_constraints_2012,dasgupta_low_2021,Chakraborty_2024,Chakraborty_low_mass_nakedsingularity2024,AHA_stall_ang_2025}, thereby upsetting the hydrostatic equilibrium. Here, we look for conditions under which, despite the EBH being present, the equilibrium condition can be restored and the accretion of host matter by EBH can be significantly slowed, due to the magnetic field of the host. 
 We assume that the magnetic field is approximately uniform at the core of the host star. This is justified because the size of a newly formed EBH is much smaller than the characteristic size of the host star \cite{Chakraborty_2024,Chakraborty_low_mass_nakedsingularity2024,AHA_stall_ang_2025}, so the EBH is immersed in the local core magnetic field, which is expected to be nearly uniform over the length scales relevant to the EBH near the centre of a spherical fluid configuration \cite{Fujisawa_2013}. As accretion proceeds, inflowing matter can advect magnetic flux toward the EBH, potentially enhancing the local magnetic field strength near the EBH without significantly modifying the large-scale core field structure.
At leading order, we therefore focus only on the dominant radial stress balance between magnetic pressure, gravitational forcing, and residual matter pressure, which motivates the pressure balance condition introduced below.

For the accretion to slow down significantly, we should have, 
\begin{equation}
    \frac{B^2}{8\,\pi}\,= \,\frac{G\,M\,\rho}{ R}+P_{\rm h}
    \label{press}
\end{equation}
where $B$ is the strength of the magnetic field in the
host’s core; $G$ is Newton’s gravitational constant; $\rho$ is the average density of the host; $M$ is the mass of the
EBH at any instance of temporary equilibrium; $R$ is the radius of equilibrium where pressures balance and cancel each other exactly. Note that Eq. (\ref{press}) is intended as a phenomenological, leading-order scaling relation representing the local pressure balance at the onset of magnetic arrest. In a simplified quasi-steady framework, the radial Euler equation contains contributions from gravitational force, residual matter pressure gradients, and magnetic stress gradients. Estimating these terms at the level of order-of-magnitude scaling leads naturally to the local balance expressed in Eq. (1). The relation should therefore be interpreted as an order-of-magnitude arrest criterion rather than as an exact global hydrodynamical or GRMHD solution. The first term on the right-hand side of  Eq. \eqref{press} is due to the gravitation of the EBH acting on the host matter, in other words, the gravitational energy density or the gravitational pressure and the second term ($P_{\rm h}$) comes from the residual pressure of the host matter. 
$P_{\rm h}$ represents the pressure of all other possible origins, such as the degeneracy pressure, thermal pressure, etc. This term might be initially several orders of magnitude less than the gravitational pressure represented by the first term on the right-hand side of  Eq. \eqref{press}, in case of accretion flow onto the EBH, but becomes important when magnetic arrest of accretion begins.

The dominant pressure that adds to the gravitational pressure in compact stars is the degeneracy pressure of matter. The residual matter pressure is therefore typically dominated by degeneracy pressure. We can consider a polytropic equation of state (EOS) $P = K \rho^{\Gamma}$ where $\Gamma$ is the effective polytropic index, and obtain the residual pressure $P_{\rm h}$ by plugging the central density of the host for $\rho$. The magnitude of $P_{\rm h}$ depends on the host object and roughly in the same order as the central pressure. For example, the central pressure in WDs lies in the range $10^{19}-10^{26}~\mathrm{dyne~cm^{-2}}$ for a polytropic EOS with $\Gamma=5/3$ (non-relativistic limit at lower densities) or $\Gamma=4/3$ (ultra-relativistic limit at higher densities) \cite{Bhattacharya_2022}. In NS cores, it typically lies in the range $\sim 10^{33}-10^{36}~\mathrm{dyne~cm^{-2}}$, for example in piecewise polytropic models of the form $P(\rho)=K_i\rho^{\Gamma_i}$ with $1<\Gamma_3<5$ \cite{J_Read_2009_EOS}.

Now, writing $R (:=rGM/c^2)$ in terms of the gravitational radius ($GM/c^2$) of EBH, we can express  Eq. \eqref{press} as $ B^2/8\,\pi\,= \,\rho c^2/r+P_{\rm h}$,
which can be seen to be a comparison between energy densities of magnetic field and of the host matter. Here, $r$ is the magnetospheric radius in gravitational radius units. 
Now, from  Eq. \eqref{press} we can write,
\begin{equation}
    R=\frac{8\pi\, G M \rho}{(B^2-8\pi P_{\rm h})}.
    \label{radius}
\end{equation}
 Eq. \eqref{radius} gives the radius where the magnetic pressure and gravitational pressures balance each other. This may not happen immediately after the formation of EBH as the matter density and pressure at the core exceed the magnetic pressure. 
However, as the EBH grows by accreting matter, the matter density and pressures decrease and the magnetic field does not decrease as rapidly since, by the no-hair theorems \cite{Israel_1967,israel_event_1968} a BH cannot support an intrinsic magnetic field and so advection of the magnetic flux is the only channel for the field to decrease.
This however is a slower process in a small region around EBH since the currents that sustain the field are much larger than the horizon. 
Thus, at some point in the growth of EBH, magnetic pressure may become comparable to gravitational pressure and other pressures acting on host matter and attain an equilibrium state. 
In equilibrium, the EBH cannot grow further significantly. 
Thus, the final mass ($M_{\rm f}$) of EBH is obtained in terms of $M_0$  as
\begin{equation}
   M_{\rm f}= M_0\,S(\beta)
    \label{FMass}
\end{equation}
where, 
\begin{equation}
S(\beta)=1+\beta+3\beta^2+12\beta^3+55\beta^4+....
\label{sb}
\end{equation}
(see \ref{appendix} for derivation) with \\
$\beta =\left(2048\,G^{3}\pi^4\rho^4\right)M_0^2/\left(3\,(B^2-8\pi P_{\rm h})^{3}\right)$. 
It is quite apparent that if $0<\beta \leq  4/27$, the series converges and hence accretion is stalled. 
For $\beta>4/27$, the function $S(\beta)$ diverges, and hence the final mass approaches the host mass, and the host is therefore transmuted into a BH. 
At the critical value $\beta=4/27$, $S(\beta)=3/2$. Therefore, the final mass of EBH, in case accretion is stalled, can at most be $M_{\rm f}=3/2\, M_0$. 
We call this phenomenon of magnetically stalled accretion, which in turn stalls the transmutation of the host, the  Magnetically Arrested Transmutation (MAT).
Here, $\beta$ is called the MAT parameter, since its numerical value determines the MAT mechanism. It is important to note here that while a complete halt of accretion could be unrealistic, for the sake of conceptual demonstration we say that accretion is halted. Realistically, accretion could continue, albeit at a slower rate, determined by the plasma parameters. Even the growth of EBH up to the final mass $M_{\rm f}$ is a gradual continuous process.

Considering  Eqs. (\ref{FMass}-\ref{sb}) with $\beta \ll 4/27$, the accretion time till the system attain MAT equilibrium would be

\begin{align}
    t_{\rm acc}\sim&\frac{\left(c_{\rm s}^2+v_{\rm A}^2\right)^{3/2}}{4 \pi \rho \,G^2 }\int^{M_{\rm f}}_{M_0}\frac{dM}{M^2}\\=&\frac{\left(c_{\rm s}^2+v_{\rm A}^2\right)^{3/2}}{4 \pi \rho \,G^2 } \left(\frac{1}{M_0}-\frac{1}{M_{\rm f}}\right)\approx \frac{\left(c_{\rm s}^2+v_{\rm A}^2\right)^{3/2}}{4 \pi \rho \,G^2} \ \frac{\beta}{M_0}
    \label{accrrtime}
\end{align}
where 
$v_{\rm A}$ is the Alfvén velocity.
While  Eq. \eqref{accrrtime} gives the accretion timescale, considering that for most compact stars, the higher probability will be to have $\beta\ll 4/27$ (refer to Fig. \ref{fig:BetavsB}), $t_{\rm acc}$ is negligible compared to the EBH formation timescale $t_0$ (see Sect. 2.2 of \cite{Chakraborty_2024}). 
For $M_{\rm f}\rightarrow 3/2\,M_0$, we obtain \\ $t_{\rm acc}^{\rm max} \rightarrow \left(c_{\rm s}^2+v_{\rm A}^2\right)^{3/2}/(12 \pi \rho \ G^2 M_0)$, the order of magnitude of which is almost similar to the ordinary accretion timescale, as considered in \cite{Chakraborty_2024,Chakraborty_low_mass_nakedsingularity2024}.

\section{Application to compact stars}
\label{sec2}

Compact stars with extremely strong magnetic fields--such as mWDs and magnetars--are generally slow rotators. 
However, due to the sudden collapse of the DM core to an EBH creates a cavity into which the fluid with a small angular momentum flows. This flow, induces radial and toroidal components of a radially varying magnetic field as matter falls into EBH. Such a field then exerts pressure on the matter that's falling radially inwards.
This raises the possibility of magnetic stalling, where accretion is arrested due to an equilibrium established between magnetic and gravitational pressures.

mWDs are a special kind of WDs proposed to explain super-Chandrasekhar mass progenitors of some observed over-luminous Type Ia supernovae. 
Such mWDs can have a surface magnetic field upto $10^{9}$ G \cite{Kalita2019} and a core magnetic field upto $10^{14}$ G \cite{Bhattacharya_2022}. We also consider mCVs in this context, following~\cite{ferrario_magnetic_2015,Xu_2019}, and refer to them collectively as mWDs. We assume a uniform average density for mWDs to simplify the analysis, with radii in the range $2{,}000-20{,}000$~km and masses spanning $0.16-1.8\,M_\odot$~\cite{Bhattacharya_2022}. This yields average densities of $10^4-10^8~\mathrm{g~cm^{-3}}$.

\begin{figure}
\begin{subfigure}[t]{0.45\textwidth}
    \includegraphics[width=\textwidth]{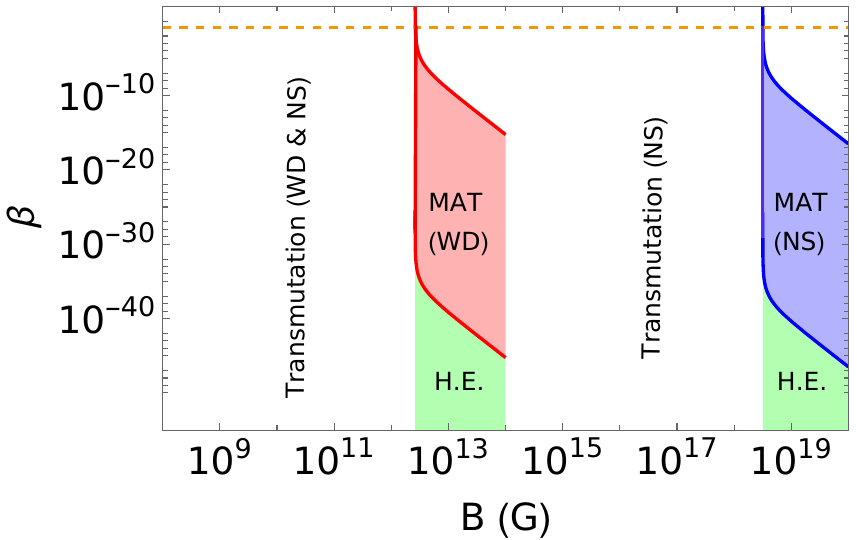}
    \caption{
        mWD: $M_{\rm mWD}=1\,M_\odot$, $\rho_{\rm avg}=2\times10^6$ g cm$^{-3}$, 
        Magnetar: $M_{\rm NS}=2 M_\odot$, $\rho_{\rm core}=8\times10^{14}$ g cm$^{-3}$ \cite{FerrerEOS}.
    }
\end{subfigure}
\hfill
\begin{subfigure}[t]{0.45\textwidth}
    \includegraphics[width=\textwidth]{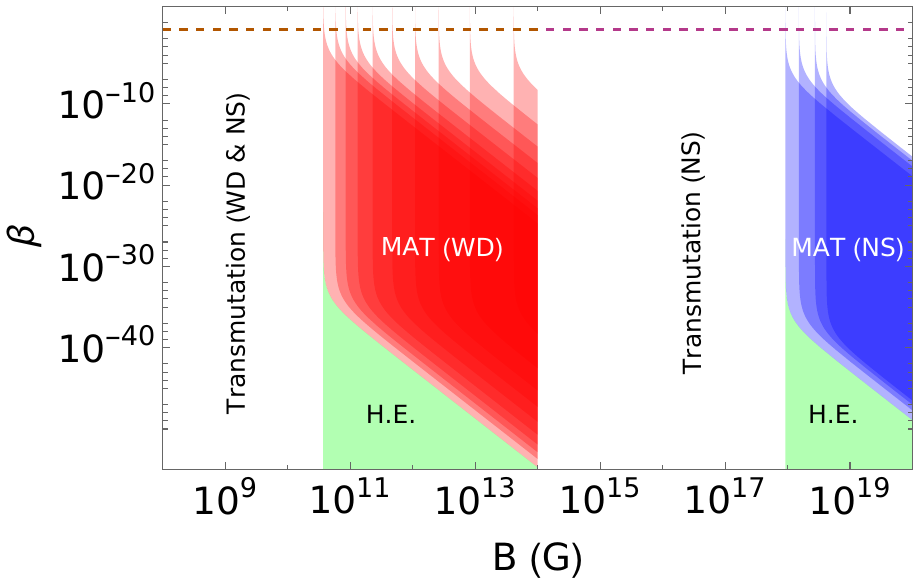}
    \caption{
        mWD: $M_{\rm mWD}\in \left[0.16, \,1.8\right]\,M_{\odot}$ \cite{Bhattacharya_2022}, 
        $\rho_{\rm avg} \in \left[10^4 ,\, 10^8\right]$ g cm$^{-3}$, 
        Magnetar: $M_{\rm NS}\in \left[1.17, \,2.35\right]\,M_{\odot}$ \cite{Martinez_2015,Romani_2022}, 
        $\rho_{\rm core} \in \left[2.8\times10^{14} ,\, 3\times10^{15}\right]$ g cm$^{-3}$.
    }
\end{subfigure}

\caption{\label{fig:BetavsB} Plot of the dimensionless MAT parameter $\beta :=\left(2048\,G^{3}\pi^4\rho^4\right)M_0^2/\left(3\,(B^2-8\pi P_{\rm h})^{3}\right)$ as a function of the core magnetic field $B$, for the EBH mass range : $10^{15}$g $\leq M_0 \leq 10^{30}$ g formed at the cores of magnetized compact stars. 
Panel (a): Light red (light blue) shaded region shows the range of $B$ for which $\beta \leq 4/27$ (dashed line), ensuring suppression of EBH growth via the MAT mechanism inside a WD (NS) of mass $1M_\odot$ ($2M_\odot$). The boundaries of each shaded region correspond to $M_0=10^{15}$~g (lower) and $10^{30}$~g (upper). Below the lower curves (i.e., the green shaded regions), Hawking Evaporation (H.E.) of the EBH ensues, and the host does
not undergo transmutation, as the EBH eventually disappears, aided by the MAT mechanism.
Panel (b): Same as (a), but for a range of WD (NS) masses and densities. Left of the shaded region, $B$ is insufficient to stall the growth of EBH and to the right, $B$ is unphysical since the magnetic energy density exceeds the energy density and pressure of the matter. 
This figure shows that the WDs (NSs) with a core magnetic field in the lower range than the shaded region transmute into BHs, while those with fields within the horizontal extent of the shaded region avoid transmutation due to the MAT mechanism or Hawking evaporation.
Note that, although H.E. region for lowest mass mWD is shown exactly in Panel (b), for mWD of higher masses, the H.E. region slightly overlaps with the red-shaded region, since the lower curve (corresponding to $M_0=10^{15}$ g) for those mWD masses is at a slightly greater value of $\beta$ than those with lower masses.}
\end{figure}

For an EBH formed in such a host, accretion halts when the final mass remains less than the host mass. This occurs when $0 < \beta \leq 4/27$ (see  Eq.~\ref{FMass}). In Panel (a) of Fig. \ref{fig:BetavsB}, the light-red shaded region shows the allowed values in the $\beta-B$ space for a fixed host mass and density. 
The lower and upper curves correspond to initial EBH masses of $10^{15}$~g and $10^{30}$~g, respectively. The lower bound is set by significant Hawking evaporation below $10^{15}$~g~\cite{C_Chakraborty_22_gravmag}, while the upper bound, on the other hand, is conservatively chosen so that the EBH formation time 
($t_0^{\rm f}$ : \cite{Chakraborty_2024}) does not exceed the age of the host (see Fig. 1 in conjunction with Fig. 2 of \cite{Chakraborty_low_mass_nakedsingularity2024}). For all the values of initial mass between these two bounds, the value that $\beta$ takes lies within the light-red shaded region. 
For host mass $M_{\rm mWD}$ in the range $0.16-1.8\,M_\odot$ and the corresponding average density, as given in Table 3 of \cite{Bhattacharya_2022}, we can similarly plot (see the light-red shaded region in Panel (b) of Fig.~\ref{fig:BetavsB}) the values that $\beta$ takes, with the same lower and upper bounds on initial mass, against $B$. 
This reveals that a wider range of magnetic field allows for a MAT scenario, with lower host mass showing possibilities for MAT at a core magnetic field strength of $\sim 10^{10}$ G. To the left side of the shaded region, the magnetic field strength is insufficient to restore equilibrium, and to the right side, the magnetic field strength is unphysically high. 
Also note that the magnetic field is unphysically high above $\sim 10^{13}$ G for some of the mWDs at the lower end of the mass range. Below the light red/blue shaded region (i.e., within the green shaded regions), the initial mass $M_0<10^{15}$ g  leads to a relatively high mass loss rate due to Hawking radiation. 
Given that  $\beta \ll 1$ for such $M_0$ values, the MAT mechanism becomes operative. Consequently, the evaporation of the EBH proceeds more rapidly than it would in the absence of MAT, since there is no compensating mass accretion.  In this scenario, the host object does not undergo transmutation, as the EBH eventually disappears, aided by the MAT mechanism.

The strength of magnetic field at the surface of magnetars is estimated to be in the range $10^{13}\,-10^{15}$ G \cite{OriginMag}. 
Theoretical calculations show that the magnetic field strength at the core can go up to $10^{20}$ G as per calculations considering quark matter stars \cite{FerrerEOS} and $10^{17}-10^{18}$ G considering nuclear matter stars \cite{DongShapiroMag,FerrerNeutronsinMag}. The strength of the magnetic field at the core can be estimated using hydrostatic equilibrium conditions or using the scalar virial theorem \cite{DongShapiroMag} (also see \cite{frieben_equilibrium_2012} for bounds based on stability criteria). 
Hence, we consider the core magnetic field strength in magnetars to be varying between $10^{16}$ G and $10^{20}$ G \cite{FerrerNeutronsinMag}. 
The magnetic pressure on the dense nuclear matter or quark matter in the corresponding range is $\sim 10^{33}-10^{36}$ dyne cm$^{-2}$. 
These ranges of energy densities do make it possible for the condition  Eq. \eqref{press} is satisfied in a narrow range. 

Similar to mWD, MAT scenario could be possible inside magnetars
for $\beta \leq 4/27$. If one takes the same lower and upper bounds (i.e., $10^{15}$ g $\leq M_0 \leq 10^{30}$ g) on the initial mass and plots $\beta$ against magnetic field, one finds that a magnetar of mass $2M_\odot$ needs a core magnetic field above $\sim4\times10^{18}$ G to meet the conditions of MAT. This can be observed from the blue shaded region in Panel (a) of Fig. \ref{fig:BetavsB}.
The required magnetic field strength increases for a greater host mass and core density. The core density can vary in integral multiples of the nuclear density $2.8\times 10^{14}$ g cm$^{-3}$ \cite{FerrerEOS}. 
Since the core density of magnetars does not vary with host mass as drastically as it does in mWDs, the range of magnetic field for which MAT is realized is much narrower compared to the mWD case, i.e., a core magnetic field in the range $10^{18}-10^{20}$ G is necessary in magnetars to realize MAT (refer Panel (b) of Fig. \ref{fig:BetavsB}). 
From these calculations, it is apparent that there is a possibility that magnetars, formed in regions with a high density of DM, can survive without transmuting into black holes even if an EBH forms inside them, provided that they have a sufficient magnetic field to stall the growth of EBH. 
Note that the most conservative upper limits on internal magnetar magnetic fields are typically placed around $10^{17}$ G, while more optimistic estimates allow values up to $\sim 10^{18}$ G. Field strengths approaching $\sim 10^{20}$ G \cite{FerrerNeutronsinMag} are generally obtained only under assumptions of exotic phases of dense matter, and the stability of such configurations remains uncertain. Within this context, the MAT mechanism may be relevant in magnetars only in extreme cases where sufficiently strong internal magnetic fields are realized.

This could potentially offer an explanation for the observed presence of a magnetar PSR J1745-2900 \cite{Mori_2013,Pearlman_2018} in the vicinity of Sgr A*, in stark contrast with the well-known ``missing pulsars" \cite{Dexter_2014} at the GC. 
As shown in \cite{kouvaris_growth_2014,AHA_stall_ang_2025}, even though pulsars spin very rapidly, their angular momentum is not enough to stall accretion \cite{AHA_stall_ang_2025} by EBH and hence should get transmuted. 
However, the magnetars, which have non-negligible angular momentum as well as a magnetic field capable of stalling accretion by EBH, can survive longer without transmutation. 
PSR J1745–2900’s age estimates vary from $\gtrsim 37$ years \cite{Katz2020} to $\sim9000$ years \cite{Bower_2015}. If its age is at the higher end of this range ($\sim10^4$ years) and the core magnetic field is in the order of $B_c \sim 10^{18}$ G, 
there is a chance that MAT is responsible for its survival, for
the DM-nucleon scattering cross-section $\sigma \geq 10^{-44}$ cm$^2$ \cite{Chakraborty_2024} with the bosonic DM particle of mass $17 \,{\rm GeV} < m_\chi < 8\times10^4\,{\rm GeV} $ and a DM density of $\sim 7\times10^4$ GeV cm$^{-3}$ \cite{Navarro_1997} in GC. In such a scenario, MAT could explain the survival of PSR J1745-2900. 
However, a sufficiently low age or weak core magnetic field renders MAT non-essential for the survival of this magnetar.
From the measured spin-down rate, PSR J1745-2900's surface magnetic field was estimated to be $\sim10^{14}$ G \cite{Mori_2013}. It should also be noted that long-term Chandra monitoring shows a slow
decay toward a quiescent $0.3$–$10$ keV object with a luminosity of
$\sim10^{34}$ erg s$^{-1}$, while the surface hot spot temperature
has cooled from $0.9$ keV to $0.65$ keV over six years \cite{chinHu_2019,Coti_2017}. Based on these observations, one may make an estimation of core magnetic field to be much lower than $10^{18}$ G.
However, the possible presence of multiple hotspots on the surface has been suggested to be consistent with complex multipolar magnetic-field structures \cite{de_Lima_2020}. While such configurations may, in some models, be associated with stronger internal magnetic fields \cite{Mastrano_NS_deform_2013}, the connection between surface observables and any ultra-strong core magnetic field remains indirect.
Also, as per calculations with different phases of matter at core done in  \cite{FerrerEOS,DongShapiroMag,FerrerNeutronsinMag} (also discussed earlier) for the magnetized neutron stars, the mass density distribution is significantly different from a nucleon-matter core, and hence, the magnetic field energy required for equilibrium is higher \cite{FerrerEOS}. 
Furthermore, once an EBH is formed, accretion-generated luminosity could also alter this output \cite{Bellinger_2023}. Thus, a thorough investigation of these aspects would be worthwhile for PSR J1745-2900 in future work. 
While incorporating these additional effects, and obtaining a more precise estimate of PSR J1745-2900's age and core magnetic field strength, might lead to a revised conclusion for this particular magnetar, our current analysis of the MAT scenario may remain a basic framework to explain why mWDs and magnetars in GC may outlive their non-magnetic counterparts. Our scenario offers valuable insights and could play a critical role in interpreting future findings.

\section{Conclusion and discussion}
\label{sec3}

Extreme magnetic fields in mWDs and magnetars can stall the growth of EBHs. We term this mechanism MAT, analogous in spirit to the MAD model of accretion disks \cite{Narayan2003MAD}, but applicable to nearly spherical accretion inside stars.
When an EBH forms and begins accreting host matter, the built-up magnetic field around EBH may eventually exert enough pressure to balance gravity and halt further accretion. This scenario explains why mWDs and magnetars may outlive their non-magnetic counterparts, even if EBHs form in all WDs and NSs.

In contrast, ordinary pulsars cannot halt transmutation due to insufficient magnetic field, despite their rapid rotation  \cite{AHA_stall_ang_2025}. This possibility, while speculative, may provide a natural resolution to MPP, accounting for the absence of pulsars and the presence of a magnetar (PSR J1745–2900) near the GC. 
Though MAT may not prevent transmutation indefinitely, it could significantly extend magnetar lifetimes in DM-rich regions. 

For mWDs, MAT could help explain their apparent overabundance in GC relative to ordinary WDs. Although this observation perhaps is better explained in other ways, we posit that if the mechanism proposed above is true, one might actually observe more mWDs than ordinary WDs in GC. Surviving ordinary WDs may either be relatively young or rapidly rotating, enabling their EBH accretion to be stalled or slowed down via rotationally arrested transmutation (RAT) \cite{AHA_stall_ang_2025}. Improved estimates of the WD mass and magnetic field distributions in the GC would further clarify this picture. Such observable signatures may allow the MAT scenario to be tested and could help constrain the compact object population in DM-rich environments. The observational implications discussed here are necessarily qualitative. A quantitative assessment would require population-level modelling and detailed observational diagnostics, which are beyond the scope of this work. The present framework therefore identifies MAT as a physically viable channel under suitable conditions, rather than as a unique or definitive explanation.

The present model captures the leading-order competition between magnetic pressure and the combined gravitational and residual matter pressure effects responsible for MAT. The analogy with MAD is therefore intended at the level of physical mechanism rather than detailed dynamical equivalence. Unlike MAD systems, which involve disk accretion, magnetic flux transport, and non-spherical geometries, the present treatment assumes quasi-spherical accretion within the interior of a compact star. Key ingredients of fully developed MAD states, including magnetic tension, field topology, and time-dependent MHD effects, are not included here. In realistic compact-star interiors, density gradients, relativistic effects, and dynamical MHD processes are expected to smooth the sharp transition predicted by the present stepwise accretion model. The resulting critical value of $\beta$ and the associated final mass scale should therefore be interpreted as model-dependent diagnostics rather than universal analytical bounds.

Our results are obtained assuming a simple pressure balance as done in \cite{Narayan2003MAD,Igumenshchev_2008} which translates to a balance of pressure-gradient forces while neglecting dynamical feedback on the background density and magnetic field. A fully GRMHD treatment would be required to capture the evolution of field geometry and pressure gradients near the EBH and represents an important direction for future work.

\appendix

\section{\label{appendix} Mathematical Formulation for Determining the Final Mass ($M_{\rm f}$) of an EBH}

The final mass of an EBH is reached when the accretion ceases entirely and an equilibrium is established between the matter pressure and the magnetic pressure. Although accretion is a continuous process, for the purpose of mathematical formulation, we model it as a stepwise increase in mass.
In the initial step, equilibrium is attained at a radius (see Eq. \ref{radius})
\begin{align}
R_0 = \frac{8\pi\, G M_0\,\rho}{B^2-8\pi P_{\rm h}}
\end{align}
corresponding to the initial mass of EBH $M_0$.
Then, as EBH accretes matter from within $R_0$, a radius $R_1$ becomes the new radius of equilibrium. 
In the next step, mass from within $R_1$ is accreted, and EBH mass increases to $M_1$, and a new radius $R_2$ becomes the equilibrium radius. 
Within the fixed-background approximation adopted here, this sequence converges to a finite asymptotic EBH mass (final mass) for subcritical values of the control parameter ($\beta$ which we shall calculate below), as successive increases in the equilibrium radius contribute progressively smaller increments to the accreted mass.
Let us now write the mass of EBH in successive steps. The mass of EBH after accreting mass within $R_0$ is

\begin{align}
    M_1=& \,M_0+\frac{4}{3}\pi\rho R_0^3 \nonumber \\
    =& M_0+\frac{2048}{3}\frac{G^3 \pi^4 \rho^4}{\left(B^2-8\pi P_{\rm h}\right)^3}  M_0^3 \nonumber\\
    =&M_0+\alpha M_0^3
    \label{a1}
\end{align}
where $\alpha=\left(2048\,G^{3}\pi^4\rho^4\right)/\left(3\,(B^2-8\pi P_{\rm h})^{3}\right)$. Note that, in the first step, the radius of EBH is ignored compared to $R_0$. For the second step, we get
\begin{align}
    M_2=& \,M_1+\frac{4}{3}\pi\rho \left(R_1^3-R_0^3\right) \nonumber \\=&\,M_1+\alpha\left(M_1^3-M_0^3\right)
    \label{a2}
\end{align}
where, $R_1 =\left(8\pi\, G M_1\,\rho\right)/(B^2-8\pi P_{\rm h})$. Using Eq. (\ref{a1}), we rewrite Eq. (\ref{a2}) as
\begin{align}
    M_2=M_0+\alpha M_1^3.
    \label{a3}
\end{align}

\begin{figure}
    \centering
    \includegraphics[scale=0.5]{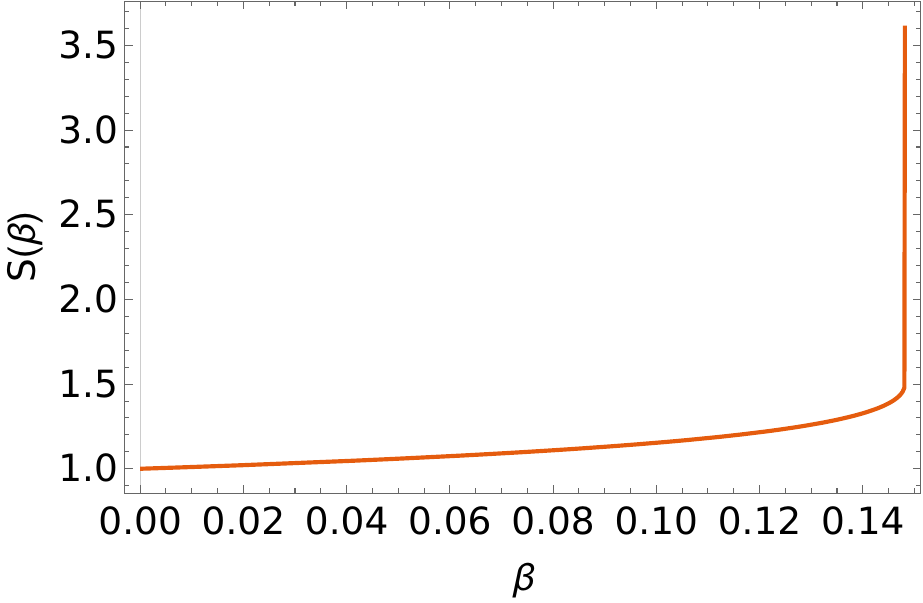}
    \caption{Plot for the nested infinite series $S(\beta)$ as a function of $\beta$. The solid orange curve shows that $S(\beta)$ has finite values for $\beta \leq 4/27$, but it diverges sharply for $\beta > 4/27$.}
    \label{fig:series}
\end{figure}

\noindent Similarly, we obtain $M_3=M_2+\alpha(M_2^3-M_1^3)=M_0+\alpha M_2^3$ and so on. Thus, for the $n$-th term one can write
\begin{align}
    M_n=M_0+\alpha M_{n-1}^3.
\end{align}
Now, by using $\beta=\alpha M_0^2$ we can rewrite the expressions for $M_1, M_2,....M_n$ as: 
\begin{align}
    M_1=&\,M_0(1+\beta) ,  \nonumber
    \\
    M_2=&\,M_0+\alpha M_1^3=M_0+\alpha M_0^3\left(1+\beta \right)^3 \nonumber
    \\ 
    =&\,M_0 \left(1+\beta \left(1+\beta\right)^3 \right) , \nonumber
    \\
    M_3=&\,M_0+\alpha M_2^3=M_0+\alpha M_0^3\left(1+\beta \left(1+\beta\right)^3\right)^3  \nonumber
    \\ 
    =&\,M_0 \left(1+\beta \left(1+\beta(1+\beta)^3\right)^3 \right) , \nonumber \\
    \vdots \nonumber\\
    M_n =&\,M_0
\left[1+\beta\left(1+\beta\left(1+\beta\left(1+\beta(.........n~\text{no. of}~\beta)^3\right)^3\right)^3\right)^3\right]. 
    \label{a4}
\end{align}
For $n \rightarrow \infty$, Eq. (\ref{a4}) reduces to 
\begin{align}
M_{\infty} :=& M_{\rm f} \nonumber
\\
=& M_0\left[1+\beta\left(1+\beta\left(1+\beta\left(1+\beta(....)^3\right)^3\right)^3\right)^3\right] \nonumber
\\
=&M_0 S(\beta)
\label{a5}
\end{align}
which is the same as Eq. \eqref{FMass}. Eq. (\ref{a5}) contains $S(\beta)$, a nested series that reduces to a functional equation of the form
\begin{equation}
    \beta \,\left[S(\beta)\right]^3-S(\beta)+1=0.
    \label{functional}
\end{equation}
Considering Eq. \eqref{functional} as a recursive relation to get the coefficients of the power series $S(\beta)=\sum_{n=0}^\infty a_n \beta^n$, we obtain a series of the form
\begin{align}
S(\beta)=& 1+\beta\left(1+\beta\left(1+\beta\left(1+\beta(....)^3\right)^3\right)^3\right)^3 \nonumber
\\
=& 1+\beta+3\beta^2+12\beta^3+55\beta^4+....
\label{a7}
\end{align}

Eq. (\ref{a7}) is expected to converge for $\beta << 1$. However, to find the radius of convergence of the series $S(\beta)$, or to find the critical value of $\beta$ above which the series $S(\beta)$ diverges, we rewrite Eq. \eqref{functional} in the form
\begin{align}
    \beta=\frac{S-1}{S^3}.
    \label{beta}
\end{align}
Differentiating Eq. \eqref{beta} with respect to $S$ and setting $d\beta/dS=0$, we obtain
\begin{align}
S=3/2.
\end{align}
Substituting for $S$ in Eq. \eqref{beta} we get $\beta=4/27 \approx 0.148$ which is the critical value of $\beta$ above which the series $S(\beta)$ diverges. The behavior of the series for $\beta \leq 4/27$ can be seen in Fig. \ref{fig:series}.

\begin{acknowledgements}
HAA acknowledges Dr. TMA Pai Ph.D. scholarship program of Manipal Academy of Higher Education (MAHE). CC acknowledges the support of MAHE. We thank the referee for
constructive comments that helped to improve the
manuscript.
\end{acknowledgements}


\bibliographystyle{spphys} 
\bibliography{MyLibrary}

\begin{thebibliography}{10}
\providecommand{\url}[1]{{#1}}
\providecommand{\urlprefix}{URL }
\expandafter\ifx\csname urlstyle\endcsname\relax
  \providecommand{\doi}[1]{DOI \discretionary{}{}{}#1}\else
  \providecommand{\doi}{DOI \discretionary{}{}{}\begingroup
  \urlstyle{rm}\Url}\fi

\bibitem{Yu_2022}
Z.L. Yu, X.J. Xu, X.D. Li, Res. Astron. Astrophys. \textbf{22}(4), 045003
  (2022).
\newblock \doi{10.1088/1674-4527/ac4e01}.
\newblock \urlprefix\url{https://dx.doi.org/10.1088/1674-4527/ac4e01}

\bibitem{Xu_2019}
X.j. Xu, Z.~Li, Z.~Zhu, Z.~Cheng, X.d. Li, Z.l. Yu, Astrophys. J.
  \textbf{882}(2), 164 (2019).
\newblock \doi{10.3847/1538-4357/ab32df}.
\newblock \urlprefix\url{https://dx.doi.org/10.3847/1538-4357/ab32df}

\bibitem{Vermette_2023}
B.~Vermette, C.~Salcedo, K.~Mori, J.~Gerber, K.D. Yoon, G.~Bridges, C.J.
  Hailey, F.~Haberl, J.~Hong, J.~Grindlay, G.~Ponti, G.~Ramsay, Astrophys. J.
  \textbf{954}(2), 138 (2023).
\newblock \doi{10.3847/1538-4357/ace90c}.
\newblock \urlprefix\url{https://dx.doi.org/10.3847/1538-4357/ace90c}

\bibitem{Kalita2019}
S.~Kalita, B.~Mukhopadhyay, Mon. Not. R. Astron. Soc. \textbf{490}(2), 2692
  (2019).
\newblock \doi{10.1093/mnras/stz2734}.
\newblock \urlprefix\url{https://doi.org/10.1093/mnras/stz2734}

\bibitem{Bhattacharya_2022}
M.~Bhattacharya, A.J. Hackett, A.~Gupta, C.A. Tout, B.~Mukhopadhyay, Astrophys.
  J. \textbf{925}(2), 133 (2022).
\newblock \doi{10.3847/1538-4357/ac450b}.
\newblock \urlprefix\url{https://dx.doi.org/10.3847/1538-4357/ac450b}

\bibitem{ferrario_magnetic_2015}
L.~Ferrario, D.~de~Martino, B.T. Gänsicke, Space Sci. Rev. \textbf{191}(1),
  111 (2015).
\newblock \doi{10.1007/s11214-015-0152-0}.
\newblock \urlprefix\url{https://doi.org/10.1007/s11214-015-0152-0}

\bibitem{Pretorius_2013}
M.L. Pretorius, C.~Knigge, A.D. Schwope, MNRAS \textbf{432}(1), 570 (2013).
\newblock \doi{10.1093/mnras/stt499}.
\newblock \urlprefix\url{https://doi.org/10.1093/mnras/stt499}

\bibitem{Pfahl_2004}
E.~Pfahl, A.~Loeb, Astrophys. J. \textbf{615}(1), 253 (2004).
\newblock \doi{10.1086/423975}.
\newblock \urlprefix\url{https://dx.doi.org/10.1086/423975}

\bibitem{Dexter_2014}
J.~Dexter, R.M. O'Leary, Astrophys. J. Letters \textbf{783}(1), L7 (2014).
\newblock \doi{10.1088/2041-8205/783/1/L7}.
\newblock \urlprefix\url{https://dx.doi.org/10.1088/2041-8205/783/1/L7}

\bibitem{Mori_2013}
K.~Mori, E.V. Gotthelf, S.~Zhang, H.~An, F.K. Baganoff, N.M. Barrière, A.M.
  Beloborodov, S.E. Boggs, F.E. Christensen, W.W. Craig, F.~Dufour, B.W.
  Grefenstette, C.J. Hailey, F.A. Harrison, J.~Hong, V.M. Kaspi, J.A. Kennea,
  K.K. Madsen, C.B. Markwardt, M.~Nynka, D.~Stern, J.A. Tomsick, W.W. Zhang,
  Astrophys. J. Lett. \textbf{770}(2), L23 (2013).
\newblock \doi{10.1088/2041-8205/770/2/L23}.
\newblock \urlprefix\url{https://dx.doi.org/10.1088/2041-8205/770/2/L23}

\bibitem{Pearlman_2018}
A.B. Pearlman, W.A. Majid, T.A. Prince, J.~Kocz, S.~Horiuchi, Astrophys. J.
  \textbf{866}(2), 160 (2018).
\newblock \doi{10.3847/1538-4357/aade4d}.
\newblock \urlprefix\url{https://dx.doi.org/10.3847/1538-4357/aade4d}

\bibitem{Johnston2006}
S.~Johnston, M.~Kramer, D.R. Lorimer, A.G. Lyne, M.~McLaughlin, B.~Klein, R.N.
  Manchester, Mon. Not. R. Astron. Soc.: Lett. \textbf{373}(1), L6 (2006).
\newblock \doi{10.1111/j.1745-3933.2006.00232.x}.
\newblock \urlprefix\url{https://doi.org/10.1111/j.1745-3933.2006.00232.x}

\bibitem{Deneva_2009}
J.S. Deneva, J.M. Cordes, T.J.W. Lazio, Astrophys. J. \textbf{702}(2), L177
  (2009).
\newblock \doi{10.1088/0004-637X/702/2/L177}.
\newblock \urlprefix\url{https://dx.doi.org/10.1088/0004-637X/702/2/L177}

\bibitem{Macquart_2010}
J.P. Macquart, N.~Kanekar, D.A. Frail, S.M. Ransom, Astrophys. J.
  \textbf{715}(2), 939 (2010).
\newblock \doi{10.1088/0004-637X/715/2/939}.
\newblock \urlprefix\url{https://dx.doi.org/10.1088/0004-637X/715/2/939}

\bibitem{Bates2011}
S.D. Bates, S.~Johnston, D.R. Lorimer, M.~Kramer, A.~Possenti, M.~Burgay,
  B.~Stappers, M.J. Keith, A.~Lyne, M.~Bailes, M.A. McLaughlin, J.T. O'Brien,
  G.~Hobbs, Mon. Not. R. Astron. Soc. \textbf{411}(3), 1575 (2011).
\newblock \doi{10.1111/j.1365-2966.2010.17790.x}.
\newblock \urlprefix\url{https://doi.org/10.1111/j.1365-2966.2010.17790.x}

\bibitem{ferrer_axion-polaritons_2024}
E.J. Ferrer, V.d.l. Incera, Eur. Phys. J. C \textbf{84}(2), 133 (2024).
\newblock \doi{10.1140/epjc/s10052-024-12486-2}.
\newblock \urlprefix\url{https://doi.org/10.1140/epjc/s10052-024-12486-2}

\bibitem{Fuller_PBH_r-process}
G.M. Fuller, A.~Kusenko, V.~Takhistov, Phys. Rev. Lett. \textbf{119}, 061101
  (2017).
\newblock \doi{10.1103/PhysRevLett.119.061101}.
\newblock
  \urlprefix\url{https://link.aps.org/doi/10.1103/PhysRevLett.119.061101}

\bibitem{fuller2015}
J.~Fuller, C.D. Ott, Mon. Not. R. Astron. Soc.: Lett. \textbf{450}(1), L71
  (2015).
\newblock \doi{10.1093/Mon. Not. R. Astron. Soc.l/slv049}.
\newblock \urlprefix\url{https://doi.org/10.1093/Mon. Not. R. Astron.
  Soc.l/slv049}

\bibitem{Caiozzo_2024}
R.~Caiozzo, G.~Bertone, F.~Kühnel, J. Cosmol. Astropart. Phys.
  \textbf{2024}(07), 091 (2024).
\newblock \doi{10.1088/1475-7516/2024/07/091}.
\newblock \urlprefix\url{https://dx.doi.org/10.1088/1475-7516/2024/07/091}

\bibitem{goldman_weakly_1989}
I.~Goldman, S.~Nussinov, Phys. Rev. D \textbf{40}(10), 3221 (1989).
\newblock \doi{10.1103/PhysRevD.40.3221}.
\newblock \urlprefix\url{https://link.aps.org/doi/10.1103/PhysRevD.40.3221}

\bibitem{kouvaris_constraining_2011}
C.~Kouvaris, P.~Tinyakov, Phys. Rev. D \textbf{83}(8), 083512 (2011).
\newblock \doi{10.1103/PhysRevD.83.083512}.
\newblock \urlprefix\url{https://link.aps.org/doi/10.1103/PhysRevD.83.083512}

\bibitem{mccullough_capture_2010}
M.~McCullough, M.~Fairbairn, Phys. Rev. D \textbf{81}(8), 083520 (2010).
\newblock \doi{10.1103/PhysRevD.81.083520}.
\newblock \urlprefix\url{https://link.aps.org/doi/10.1103/PhysRevD.81.083520}

\bibitem{mcdermott_constraints_2012}
S.D. McDermott, H.B. Yu, K.M. Zurek, Phys. Rev. D \textbf{85}(2), 023519
  (2012).
\newblock \doi{10.1103/PhysRevD.85.023519}.
\newblock \urlprefix\url{https://link.aps.org/doi/10.1103/PhysRevD.85.023519}

\bibitem{bramante_detecting_2014}
J.~Bramante, T.~Linden, Phys. Rev. Lett. \textbf{113}(19), 191301 (2014).
\newblock \doi{10.1103/PhysRevLett.113.191301}.
\newblock
  \urlprefix\url{https://link.aps.org/doi/10.1103/PhysRevLett.113.191301}

\bibitem{acevedo_supernovae_2019}
J.F. Acevedo, J.~Bramante, Phys. Rev. D \textbf{100}(4), 043020 (2019).
\newblock \doi{10.1103/PhysRevD.100.043020}.
\newblock \urlprefix\url{https://link.aps.org/doi/10.1103/PhysRevD.100.043020}

\bibitem{dasgupta_low_2021}
B.~Dasgupta, R.~Laha, A.~Ray, Phys. Rev. Lett. \textbf{126}(14), 141105 (2021).
\newblock \doi{10.1103/PhysRevLett.126.141105}.
\newblock
  \urlprefix\url{https://link.aps.org/doi/10.1103/PhysRevLett.126.141105}

\bibitem{Chakraborty_2024}
C.~Chakraborty, S.~Bhattacharyya, J. Cosmol. Astropart. Phys.
  \textbf{2024}(06), 007 (2024).
\newblock \doi{10.1088/1475-7516/2024/06/007}.
\newblock \urlprefix\url{https://dx.doi.org/10.1088/1475-7516/2024/06/007}

\bibitem{Chakraborty_low_mass_nakedsingularity2024}
C.~Chakraborty, S.~Bhattacharyya, P.S. Joshi, J. Cosmol. Astropart. Phys.
  \textbf{2024}(07), 053 (2024).
\newblock \doi{10.1088/1475-7516/2024/07/053}.
\newblock \urlprefix\url{https://dx.doi.org/10.1088/1475-7516/2024/07/053}

\bibitem{Narayan2003MAD}
R.~Narayan, I.V. Igumenshchev, M.A. Abramowicz, Publ. Astron. Soc. Jpn.
  \textbf{55}(6), L69 (2003).
\newblock \doi{10.1093/pasj/55.6.L69}.
\newblock \urlprefix\url{https://doi.org/10.1093/pasj/55.6.L69}

\bibitem{Igumenshchev_2008}
I.V. Igumenshchev, Astrophys. J. \textbf{677}(1), 317 (2008).
\newblock \doi{10.1086/529025}.
\newblock \urlprefix\url{https://dx.doi.org/10.1086/529025}

\bibitem{AHA_stall_ang_2025}
H.A. Adarsha, C.~Chakraborty, S.~Bhattacharyya, Phys. Rev. D \textbf{111},
  103033 (2025).
\newblock \doi{10.1103/PhysRevD.111.103033}.
\newblock \urlprefix\url{https://link.aps.org/doi/10.1103/PhysRevD.111.103033}

\bibitem{Fujisawa_2013}
K.~Fujisawa, Proceedings of the International Astronomical Union
  \textbf{9}(S302), 423–426 (2013).
\newblock \doi{10.1017/S1743921314002658}

\bibitem{J_Read_2009_EOS}
J.S. Read, B.D. Lackey, B.J. Owen, J.L. Friedman, Phys. Rev. D \textbf{79},
  124032 (2009).
\newblock \doi{10.1103/PhysRevD.79.124032}.
\newblock \urlprefix\url{https://link.aps.org/doi/10.1103/PhysRevD.79.124032}

\bibitem{Israel_1967}
W.~Israel, Phys. Rev. \textbf{164}, 1776 (1967).
\newblock \doi{10.1103/PhysRev.164.1776}.
\newblock \urlprefix\url{https://link.aps.org/doi/10.1103/PhysRev.164.1776}

\bibitem{israel_event_1968}
W.~Israel, Commun. Math. Phys. \textbf{8}(3), 245 (1968).
\newblock \doi{10.1007/BF01645859}.
\newblock \urlprefix\url{https://doi.org/10.1007/BF01645859}

\bibitem{FerrerEOS}
E.J. Ferrer, V.~de~la Incera, J.P. Keith, I.~Portillo, P.L. Springsteen, Phys.
  Rev. C \textbf{82}, 065802 (2010).
\newblock \doi{10.1103/PhysRevC.82.065802}.
\newblock \urlprefix\url{https://link.aps.org/doi/10.1103/PhysRevC.82.065802}

\bibitem{Martinez_2015}
J.G. Martinez, K.~Stovall, P.C.C. Freire, J.S. Deneva, F.A. Jenet, M.A.
  McLaughlin, M.~Bagchi, S.D. Bates, A.~Ridolfi, Astrophys. J. \textbf{812}(2),
  143 (2015).
\newblock \doi{10.1088/0004-637X/812/2/143}.
\newblock \urlprefix\url{https://dx.doi.org/10.1088/0004-637X/812/2/143}

\bibitem{Romani_2022}
R.W. Romani, D.~Kandel, A.V. Filippenko, T.G. Brink, W.~Zheng, Astrophys. J.
  Lett. \textbf{934}(2), L17 (2022).
\newblock \doi{10.3847/2041-8213/ac8007}.
\newblock \urlprefix\url{https://dx.doi.org/10.3847/2041-8213/ac8007}

\bibitem{C_Chakraborty_22_gravmag}
C.~Chakraborty, S.~Bhattacharyya, Phys. Rev. D \textbf{106}, 103028 (2022).
\newblock \doi{10.1103/PhysRevD.106.103028}.
\newblock \urlprefix\url{https://link.aps.org/doi/10.1103/PhysRevD.106.103028}

\bibitem{OriginMag}
C.~{Thompson}, R.C. {Duncan}, Astrophy. J. \textbf{408}, 194 (1993).
\newblock \doi{10.1086/172580}

\bibitem{DongShapiroMag}
D.~{Lai}, S.L. {Shapiro}, Astrophy. J. \textbf{383}, 745 (1991).
\newblock \doi{10.1086/170831}

\bibitem{FerrerNeutronsinMag}
E.J. Ferrer, A.~Hackebill, Phys. Rev. C \textbf{99}, 065803 (2019).
\newblock \doi{10.1103/PhysRevC.99.065803}.
\newblock \urlprefix\url{https://link.aps.org/doi/10.1103/PhysRevC.99.065803}

\bibitem{frieben_equilibrium_2012}
J.~Frieben, L.~Rezzolla, Monthly Notices of the Royal Astronomical Society
  \textbf{427}(4), 3406 (2012).
\newblock \doi{10.1111/j.1365-2966.2012.22027.x}.
\newblock \urlprefix\url{https://doi.org/10.1111/j.1365-2966.2012.22027.x}.
\newblock \_eprint:
  https://academic.oup.com/mnras/article-pdf/427/4/3406/2964186/427-4-3406.pdf

\bibitem{kouvaris_growth_2014}
C.~Kouvaris, P.~Tinyakov, Phys. Rev. D \textbf{90}(4), 043512 (2014).
\newblock \doi{10.1103/PhysRevD.90.043512}.
\newblock \urlprefix\url{https://link.aps.org/doi/10.1103/PhysRevD.90.043512}

\bibitem{Katz2020}
J.I. Katz, Mon. Not. R. Astron. Soc.: Lett. \textbf{501}(1), L76 (2020).
\newblock \doi{10.1093/mnrasl/slaa202}.
\newblock \urlprefix\url{https://doi.org/10.1093/mnrasl/slaa202}

\bibitem{Bower_2015}
G.C. Bower, A.~Deller, P.~Demorest, A.~Brunthaler, H.~Falcke, M.~Moscibrodzka,
  R.M. O'Leary, R.P. Eatough, M.~Kramer, K.J. Lee, L.~Spitler, G.~Desvignes,
  A.P. Rushton, S.~Doeleman, M.J. Reid, Astrophys. J. \textbf{798}(2), 120
  (2015).
\newblock \doi{10.1088/0004-637X/798/2/120}.
\newblock \urlprefix\url{https://dx.doi.org/10.1088/0004-637X/798/2/120}

\bibitem{Navarro_1997}
J.F. Navarro, C.S. Frenk, S.D.M. White, Astrophysical Journal \textbf{490}(2),
  493 (1997).
\newblock \doi{10.1086/304888}.
\newblock \urlprefix\url{https://dx.doi.org/10.1086/304888}

\bibitem{chinHu_2019}
C.P. Hu, C.Y. Ng, W.C.G. Ho, Mon. Not. R. Astron. Soc. \textbf{485}(3), 4274
  (2019).
\newblock \doi{10.1093/mnras/stz513}.
\newblock \urlprefix\url{https://doi.org/10.1093/mnras/stz513}

\bibitem{Coti_2017}
F.~Coti~Zelati, N.~Rea, R.~Turolla, J.A. Pons, A.~Papitto, P.~Esposito, G.L.
  Israel, S.~Campana, S.~Zane, A.~Tiengo, R.P. Mignani, S.~Mereghetti, F.K.
  Baganoff, D.~Haggard, G.~Ponti, D.F. Torres, A.~Borghese, J.~Elfritz, Mon.
  Not. R. Astron. Soc. \textbf{471}(2), 1819 (2017).
\newblock \doi{10.1093/mnras/stx1700}.
\newblock \urlprefix\url{https://doi.org/10.1093/mnras/stx1700}

\bibitem{de_Lima_2020}
R.C.R. de~Lima, J.G. Coelho, J.P. Pereira, C.V. Rodrigues, J.A. Rueda, The
  Astrophysical Journal \textbf{889}(2), 165 (2020).
\newblock \doi{10.3847/1538-4357/ab65f4}.
\newblock \urlprefix\url{https://dx.doi.org/10.3847/1538-4357/ab65f4}

\bibitem{Mastrano_NS_deform_2013}
A.~Mastrano, P.D. Lasky, A.~Melatos, Monthly Notices of the Royal Astronomical
  Society \textbf{434}(2), 1658 (2013).
\newblock \doi{10.1093/mnras/stt1131}.
\newblock \urlprefix\url{https://doi.org/10.1093/mnras/stt1131}

\bibitem{Bellinger_2023}
E.P. Bellinger, M.E. Caplan, T.~Ryu, D.~Bollimpalli, W.H. Ball, F.~Kühnel,
  R.~Farmer, S.E. de~Mink, J.~Christensen-Dalsgaard, The Astrophysical Journal
  \textbf{959}(2), 113 (2023).
\newblock \doi{10.3847/1538-4357/ad04de}.
\newblock \urlprefix\url{https://dx.doi.org/10.3847/1538-4357/ad04de}

\end{thebibliography}

\end{document}